\pdfoutput=1
\documentclass[aps,reprint,showpacs,twocolumn, superscriptaddress,preprintnumbers, amsmath, epsfig,floatfix]{revtex4-1}
\usepackage{graphicx}
\usepackage{amsmath}
\usepackage{amssymb}
\usepackage{amsfonts}
\usepackage{dcolumn}
\usepackage{dsfont}
\usepackage{latexsym}
\usepackage{rotating}
\usepackage{color}
\usepackage{latexsym}
\usepackage{bbm}
\usepackage{subfigure}
\usepackage{float}
\usepackage{epsfig}
\usepackage{psfrag}
\usepackage{natbib}
\usepackage{bm}
\usepackage{amsthm}
\usepackage{eucal}
\usepackage{mathrsfs}
\usepackage{url}
\usepackage{braket}
\usepackage{mathtools}
\usepackage{amsmath,amssymb}
\usepackage{epsfig}
\usepackage{amsmath,amssymb}
\usepackage{hyperref}
\usepackage{color}

\usepackage{balance}

\usepackage{hyperref}
\hypersetup{
colorlinks=true,final=true,
        linkcolor=blue,
        citecolor=blue,
        filecolor=blue,
        urlcolor=blue,
}

\begin{document}

\title{Investigation of the Electronic Structure and Spin-State Crossover in LaCoO$_3$ Using Photoemission Spectroscopy}

\author{Sayari Ghatak}
\affiliation{Institute of Physics, Sachivalaya Marg, Bhubaneswar 751005, India}
\affiliation{Homi Bhabha National Institute, Training School Complex, Anushakti Nagar, Mumbai 400094, India}

\author{Abhishek Das}
\affiliation{Institute for Solid State Physics, The University of Tokyo, Kashiwa, Chiba 277-8581, Japan}
\affiliation{Institute of Physics, Sachivalaya Marg, Bhubaneswar 751005, India}
\author{Andrei Gloskovskii}
\affiliation{Deutsches Elektronen-Synchrotron DESY, Notkestrasse 85, D-22607 Hamburg, Germany}

\author{Dinesh Topwal}
\email{dinesh.topwal@iopb.res.in}
\affiliation{Institute of Physics, Sachivalaya Marg, Bhubaneswar 751005, India}
\affiliation{Homi Bhabha National Institute, Training School Complex, Anushakti Nagar, Mumbai 400094, India}

\begin{abstract}
Photoemission spectroscopy is a powerful technique for studying electronic structure and spin-state transitions, as it reveals changes in the orbital configuration accompanying a spin-state crossover. In this report, we combine excitation-energy, temperature-, and geometry-dependent photoemission measurements to probe the electronic structure of LaCoO$_3$ across its thermally driven spin-state transition. By systematically comparing valence-band spectra across a wide photon-energy window—from surface-sensitive soft x-ray photoemission spectroscopy (SXPS) to bulk-sensitive hard x-ray photoemission spectroscopy (HAXPES)—we identify the Co-3$d$-derived feature ($\mathit{A}$) along with the O-2$p$-dominated features ($\mathit{B}$ and $\mathit{C}$), and explain their relative evolution in terms of photon-energy-dependent photo-ionization cross-section ratios. The thermally induced spin-state crossover is demonstrated using temperature-dependent SXPS valence-band spectra which show a progressive suppression of the feature $\mathit{A}$ with heating. Geometry-dependent HAXPES measurements further clarify how the signature of spin-state transition in LaCoO$_3$ is intricately linked to the orbital-selective response of $t_{2g}$ and $e_g$ states. Additionally, angular-dependent photo-ionization cross-section analysis provides a consistent description of the polarization dependence observed in HAXPES. Finally, configuration-interaction analysis of the Co 2$p$ core-level spectra reveals that LaCoO$_3$ evolves from a predominantly low-spin ground state at low temperature to a mixed low-spin-high-spin configuration at elevated temperatures, with the high-spin fraction reaching about 30\% at 400~K. The temperature evolution of the core-level line shape thus establishes Co 2$p$ photoemission as a sensitive quantitative probe of spin-state transitions in LaCoO$_3$.
\end{abstract}

\maketitle

\section{Introduction}
The perovskite oxide LaCoO$_3$, which crystallizes in a pseudo-cubic structure (space group $R\bar{3}c$), has been studied extensively for more than seven decades because of its intriguing magnetic behavior and related metal--insulator transitions~\cite{goodenough1958interpretation,jonker1966magnetic,heikes1964magnetic,blasse1965magnetic}. At low temperatures, LaCoO$_3$ is a nonmagnetic insulator with all six $3d$ electrons of Co$^{3+}$ occupying the $t_{2g}$ orbitals, forming a low-spin (LS, $t_{2g}^{6}e_{g}^{0}$, $S = 0$) configuration because the crystal-field splitting exceeds the intra-atomic exchange energy~\cite{raccah1967first,goodenough1965complex}. Upon heating, magnetic-susceptibility and transport measurements reveal that LaCoO$_3$ undergoes a gradual transformation to a paramagnetic semiconductor around 50--100~K and subsequently to a poor metal near 400--600~K \cite{blasse1965magnetic,raccah1967first}. These transitions are generally attributed to thermally driven spin-state transitions of the Co$^{3+}$ ions from LS to high-spin (HS, $S = 2$, $t_{2g}^4 e_g^2$) \cite{goodenough1958interpretation,goodenough1965complex,raccah1967first}. Despite decades of experimental and theoretical effort, the nature of the thermally excited spin state in LaCoO$_3$ remains controversial. It is still debated whether the LS state transforms to an intermediate-spin state (IS, $S = 1$, $t_{2g}^{5}e_{g}^{1}$) or to a mixed-spin (LS+HS) state in the intermediate temperature range (approximately in between 100-400K), and eventually transforms to the HS state, owing to the near degeneracy of several spin configurations—LS, IS and HS---which lie close in energy \cite{korotin1996intermediate}. Early interpretations following Goodenough’s model attributed the low-temperature phase to LS Co$^{3+}$ and the high-temperature phase to HS Co$^{3+}$, with an inhomogeneous mixture of LS and HS ions at intermediate temperatures~\cite{raccah1967first,goodenough1971metallic}. However, density-functional theory (LDA+$U$) studies proposed that strong Co~$3d$–O~$2p$ hybridization stabilizes the IS configuration with lower total energy than the HS state~\cite{korotin1996intermediate}. Such conflicts in theoretical studies regarding the spin state in the intermediate temperature range have also been reflected in experimental investigations. For example, inelastic neutron scattering and x-ray absorption spectroscopy (XAS) measurements have reasserted a mixed LS–HS picture, with a 1:1 population emerging near 500~K, supported by GGA+$U$ calculations~\cite{asai1994neutron,haverkort2006spin}. Evidence for a mixed-spin scenario in the intermediate temperature range has also been reported from electron spin resonance and magnetic circular dichroism and other experiments~\cite{radaelli2002structural,abbate1993electronic,haverkort2006spin,rao2004spin}. On the other hand, several Raman, XAS, electron-energy-loss, and photoemission studies have been interpreted as evidence for an IS scenario~\cite{klie2007direct,saitoh1997electronic,vanko2006temperature,noguchi2002evidence,zobel2002evidence}. The IS model is often justified by invoking strong $p$--$d$ covalency that produces charge-transfer configurations ($d^7\underline{L}$, $d^8\underline{L}^2$) and  relate the thermally induced changes in the spectral features to the structural and electronic transitions ~\cite{korotin1996intermediate}. Resonant inelastic x-ray scattering (RIXS) results further suggest that the lowest excited state in LaCoO$_3$ is predominantly HS, with negligible IS contribution~\cite{tomiyasu2017coulomb}. The key challenge in resolving the spin-state evolution of LaCoO$_3$ arises from the fact that different experimental probes often provide apparently conflicting pictures due to their distinct sensitivities to the spin state. While traditional techniques such as neutron scattering, magnetic measurements and others offer valuable insights, they yield only indirect information about the occupied electronic states. In this context, a spectroscopic method capable of directly resolving the occupied electronic structure is essential for developing a coherent understanding of the underlying electronic behavior. Photoemission spectroscopy (PES), which directly probes the Co~$3d$ states that underpin the spin-state crossover in LaCoO$_3$, therefore provides a powerful and complementary approach. Indeed, recent bulk-sensitive HAXPES studies by Takegami \textit{et al.}~\cite{takegami2023paramagnetic} have demonstrated that the paramagnetic phase of LaCoO$_3$ consists of a highly inhomogeneous mixture of LS and HS Co$^{3+}$ ions. The limited number of such photoemission studies motivates a systematic re-examination of the spin-state transition in LaCoO$_3$ using PES across different photon-energy and temperature regimes. The purpose of this study is to  explore the photon-energy, geometry, and temperature dependence of the electronic structure by combining SXPS and HAXPES, thereby providing a more comprehensive understanding of the evolution of the Co$^{3+}$ spin state in LaCoO$_3$. In this work, we present a systematic investigation of the electronic structure of LaCoO$_3$ over a wide range of photon energies and temperatures to elucidate the spin-state transition. Clear signatures of the spin-state crossover are manifested by the suppression of the Co~$3d$ peak near the Fermi level in the valence band and the emergence of characteristic features in the core-level photoemission spectra. By analyzing the Co~$2p$ core-level photoemission spectra using a full multiplet configuration-interaction (CI) model, we provide a quantitative description of these changes. Our results reveal that LaCoO$_3$ evolves from a predominantly LS ground state at low temperature to an inhomogeneous mixed-spin configuration composed of LS and HS Co$^{3+}$ ions at higher temperatures.

\section{Methods}
\subsection{Sample Synthesis and Characterization}
Polycrystalline LaCoO$_3$  sample was synthesized by citric gel method \cite{zhang1987preparation, taguchi1997synthesis, natile2007lacoo3}. Aqueous solutions of stoichiometric La(NO$_3$)$_3\cdot$6H$_2$O (Alfa Aeser, 99.99\%) and Co(NO$_3$)$_2\cdot$6H$_2$O (CDH, 99\%) were prepared in a 1:1 ratio. Citric acid (Anhydrous, CDH, 99\%) was added as a complexing agent in twice the molar amount of the precursors. The solution was evaporated at 80~$^\circ$C under magnetic stirring, yielding a jelly-like substance that was dried at 90~$^\circ$C, crushed, and calcined in air at 700~$^\circ$C for 2~h. The resulting well grounded powder was pressed into pellets, annealed at 1250~$^\circ$C for 24~h, and cooled to room temperature. Room-temperature X-ray diffraction (XRD) measurements using a Cu--K$\alpha$  source reveal that the synthesized sample crystallizes in a rhombohedral $R\overline{3}c$ symmetry, consistent with previous reports
\cite{goodenough1965complex,thornton1986neutron,raccah1967first}.

\subsection{Photoemission Spectroscopy (PES)}

The HAXPES measurements were performed at the P22 beamline of the PETRA~III synchrotron light source (DESY, Germany) at different temperatures with an incident photon energy of 6~keV. Data were acquired in two geometries, namely parallel (P-polarization) and perpendicular (S-polarization) \cite{takegami2019valence}. 
SXPS was  performed using a Kratos Analytical Axis Supra+ spectrometer with monochromatic Al K$\alpha$ (1486.6~eV) photon source. Valence-band and Co~2$p$ core-level spectra of LaCoO$_3$ were recorded at different temperatures under a base pressure of $\sim$10$^{-10}$~Torr. The wide-scan x-ray photoelectron spectrum (survey scan) contained only signals corresponding to La, Co, and O, confirming the absence of extraneous elements.

\subsection{Cluster-Model Calculations}

The Co~2$p$ core-level spectra were calculated using charge-transfer multiplet cluster-model calculations using \textsc{Quanty} code\cite{haverkort2014bands,lu2014efficient,haverkort2012multiplet}. The system was modeled as a CoO$_6$ cluster in octahedral ($O_h$) symmetry, and the electronic parameters were tuned to reproduce the experimental spectra. The ground state was expanded in terms of $3d^6$, $3d^7\underline{L}$, and $3d^8\underline{L}^2$ configurations, where $\underline{L}$ indicates a hole in the oxygen ligand band.    

The model includes the on-site Coulomb interaction between $3d$ electrons ($U_{dd}$), the $2p$–$3d$ Coulomb interaction ($U_{pd}$), the crystal-field splitting ($10Dq$), and the charge-transfer energy ($\Delta$), defined as the energy required to transfer an electron from the ligand ($2p$) orbital to the metal ($3d$) orbital. The $p$–$d$ hybridization strengths were parameterized as $V_{eg}$ and $V_{t_{2g}}$, with $V_{t_{2g}} = \tfrac{1}{2}V_{eg}$.  
The parameters used are closed to earlier work and the spin-state is stabilized by choosing different crystal field values.

The $3d$–$3d$ multiplet interactions were described by the Slater integrals $F^2_{dd}$ and $F^4_{dd}$, and the $2p$–$3d$ interactions by $F^2_{pd}$ and $G^1_{pd}$, whose values were scaled to 90\% of the Hartree–Fock values to account for intra-atomic configuration interaction effects. Spin–orbit couplings for both $2p$ and $3d$ shells were included explicitly to match the experimentally observe splitting \cite{haverkort2005spinorbitaldegreesfreedom}.  To compare the calculated spectra with experiment, the discrete final states were broadened using a Gaussian convolution (0.30 eV FWHM) to account for the instrumental energy resolution. A global Lorentzian broadening of 1.1 eV was applied to represent the overall core-hole lifetime. In addition, a small multiplet-dependent Lorentzian broadening was applied to individual final-state peaks (2$p_{3/2}$ and 2$p_{1/2}$)—0.27–0.29 eV for the LS spectrum and 0.47–0.49 eV for the HS spectrum—to account for the intrinsic lifetime differences of the corresponding multiplet groups. This two-step broadening procedure yields an accurate reproduction of both the main-line asymmetry and the satellite intensity in the experimental Co 2$p$ spectra.

\section{Results and Discussion}

\subsection{Valance Band Photo-Emission Spectra}

\begin{figure}[t]
    \centering
    \includegraphics[scale=0.4]{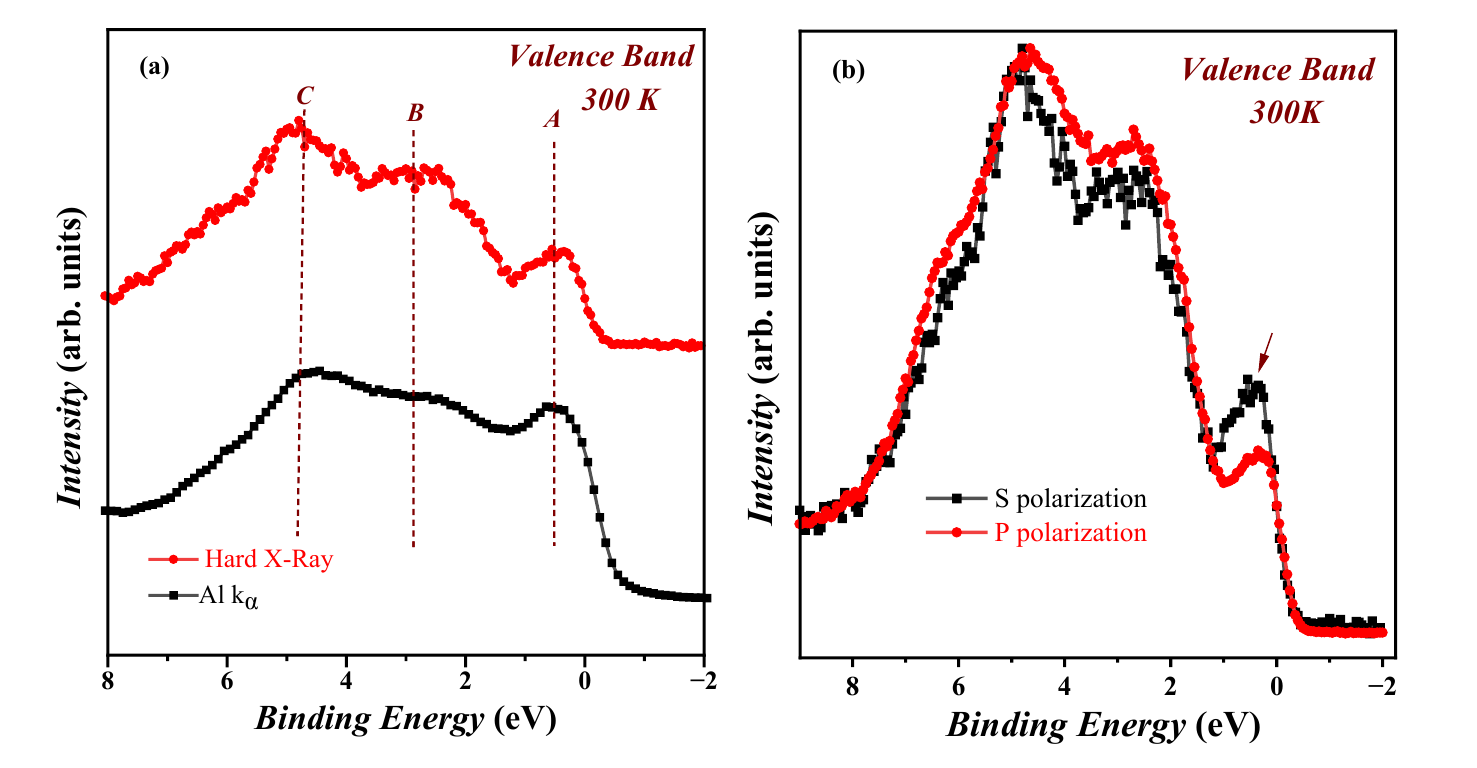}
    \caption{(a) Valence-band photoemission spectra of LaCoO$_3$ recorded with Al~K$\alpha$ (SXPS, 1486.6~eV) and hard X-rays (HAXPES, 6000~eV) at room temperature.
    (b) Comparison of HAXPES valence-band spectra of LaCoO$_3$ recorded with S and P polarizations at room temperature.}
    \label{fig:valence1}
\end{figure}

To investigate the distinct features of the electronic structure of LaCoO$_3$ over a wide range of photon energies, we first present the valence-band (VB) photoelectron spectra of LaCoO$_3$ recorded at room temperature using two different photon energy sources, Al~K$\alpha$ (1486.6~eV) (SXPS) and hard x rays (6~keV) (HAXPES) in Fig.~\ref{fig:valence1}(a). Three main features can be identified in the VB spectra at binding energies of about 1.0~, 3.0~, and 5.0~~eV marked by $\mathit{A}$, $\mathit{B}$ and $\mathit{C}$ respectively. These features originate from the hybridization of Co 3d and O 2p states. Based on the tabulated photo-ionization cross sections \cite{yeh1985atomic,yeh1993atomic}, the feature $\mathit{A}$ near 1~eV mainly is dominated by contribution from Co~3$d$ , while features $\mathit{B}$ and $\mathit{C}$ arise primarily from O~2$p$ emissions in  SXPS. This qualitative assignment agrees well with simulated valence-band spectra obtained from LDA calculations~\cite{takegami2019valence} and other studies \cite{barman1994photoelectron,richter1980ultraviolet}. A comparison between the SXPS and HAXPES valence-band spectra reveals that the relative intensity of feature $\mathit{A}$ is considerably smaller than that of features $\mathit{B}$ and $\mathit{C}$ in HAXPES compared with SXPS, despite the comparable Co~$3d$/O~$2p$ photo ionization cross-section ratios at Al~K$\alpha$ (9.44) and hard x-ray (8.60) photon energies \cite{yeh1985atomic,yeh1993atomic},  (Table~\ref{tab:subshell_cs}). The geometry dependence of the HAXPES valence-band spectra measured at 300~K is depicted in Fig.\ref{fig:valence1}(b). Interestingly, feature~$\mathit{A}$ in $S$-polarization is enhanced relative to $P$-polarization (the spectra are normalized to their total integrated intensities), which contradicts the polarization-dependent photo-ionization cross sections listed in Table~\ref{tab:subshell_cs}. The tabulated values at a photon energy of 6~keV (Table~\ref{tab:subshell_cs}), which predict a decrease in feature~$\mathit{A}$ relative to features~$\mathit{B}$ and~$\mathit{C}$ when going from $P$ to $S$ geometry, are interpolated from several literature reports \cite{trzhaskovskaya2001photoelectron,trzhaskovskaya2002photoelectron,trzhaskovskaya2006non}. To understand this apparent discrepancy, Takegami \textit{et al.}~\cite{takegami2019valence} performed LDA-based calculations and showed that La~5$p$ states contribute significantly to the HAXPES valence-band spectra.

\begin{figure}[t]
    \centering
    \includegraphics[scale=0.4]{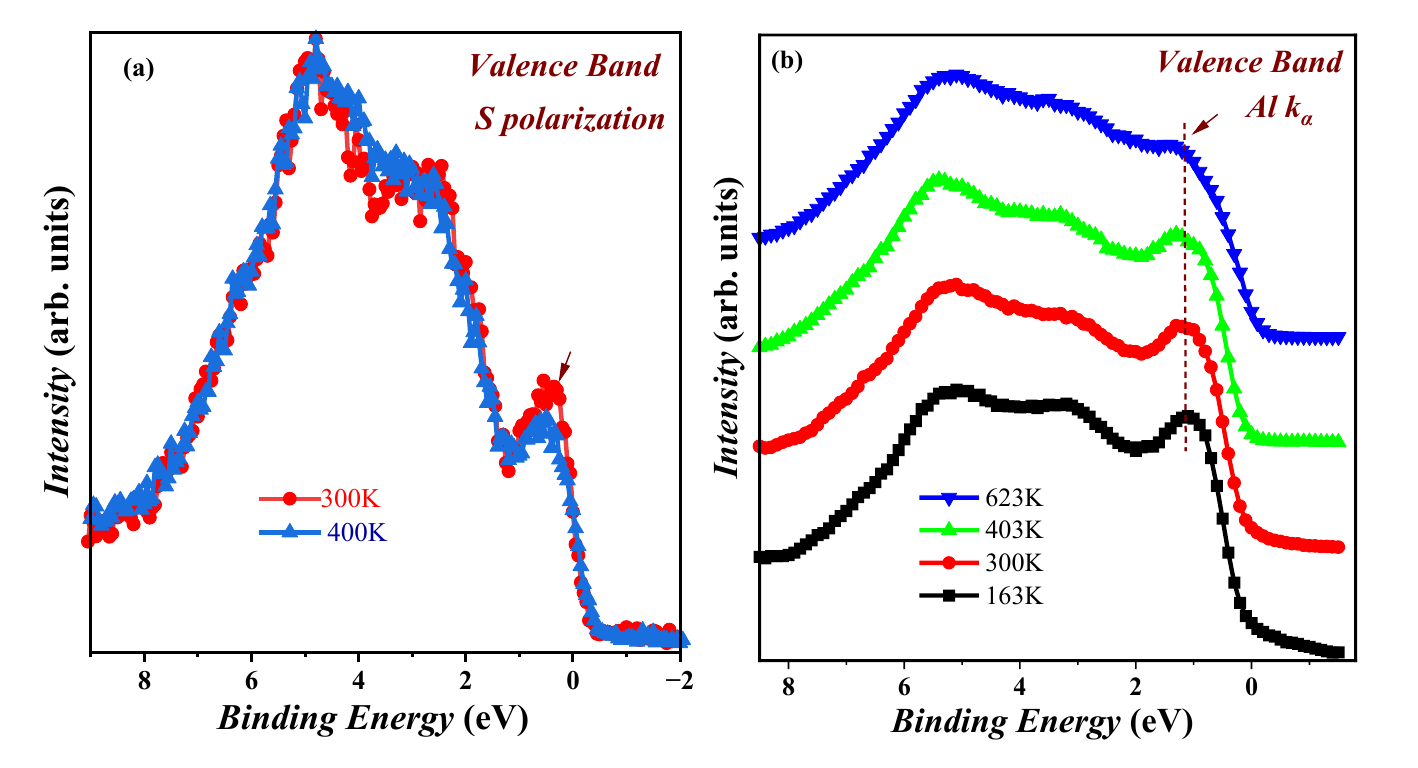}
    \caption{(a) Temperature-dependent HAXPES valence-band spectra of LaCoO$_3$ obtained with S polarization.
    (b) Temperature-dependent SXPS valence-band spectra of LaCoO$_3$ measured with an Al~K$\alpha$ source.}
    \label{fig:valence2}
\end{figure}

\renewcommand{\thetable}{\Roman{table}}
\begin{table}[htbp]
\caption{Subshell photoionization cross sections and dipole angular distribution parameters at $h\nu = 6.0$~keV, deduced from Refs.~\cite{trzhaskovskaya2001photoelectron,trzhaskovskaya2002photoelectron,trzhaskovskaya2006non}. Cross sections are normalized per electron in each subshell. Parallel ($\sigma_{\parallel}$) and perpendicular ($\sigma_{\perp}$) geometries were obtained using $\sigma[1+\beta(\tfrac14+\tfrac34\cos2\theta)]$ with $\theta = 0^\circ$ and $90^\circ$, respectively. All cross-section values are given in units of $10^{-3}$~kb}
\label{tab:subshell_cs}

\centering
\setlength{\tabcolsep}{6pt}
\begin{ruledtabular}
\begin{tabular}{l l r r r r}
Element & Subshell & $\sigma$/e & $\beta$ & $\sigma_{\parallel}$ & $\sigma_{\perp}$ \\
        &          &  &  &  &  \\
\hline
Co & $3d_{3/2}$ & 1.393 & 0.392 & 1.939 & 1.120 \\
La & $5p_{3/2}$ & 77.073 & 1.588 & 199.526 & 15.846 \\
O  & $2p_{3/2}$ & 0.306 & 0.131 & 0.346 & 0.285 \\
\end{tabular}
\end{ruledtabular}
\end{table}

 Importantly, they showed that the use of S-polarized light suppresses the La~5$p$ contribution much more strongly than the Co~3$d$ signal. Quantitatively, the relative intensity ratio La~5$p$/Co~3$d$ decreases from $\sim$43.5 in P polarization to $\sim$6.5 in S polarization, corresponding to a reduction by nearly a factor of seven. Similarly, the La~5$p$/O~2$p$ ratio drops from $\sim$597 to $\sim$62 upon switching from P to S polarization. This strong polarization dependence of the La~5$p$ contribution naturally explains the relative enhancement of Co~3$d$ features observed in the S-polarized HAXPES spectra [Figs.~\ref{fig:valence1}(a) and \ref{fig:valence1}(b)]. Takegami \textit{et al.}~\cite{takegami2019valence} also highlighted  that  the La~5$p$ contribution to the near-$E_F$ valence-band region (feature~$\mathit{A}$) in SXPS, is much smaller compared to HAXPES, suggesting that Co~3$d$ states dominate the spectral weight in this energy range. To investigate the temperature evolution of the valence-band features to probe the spin-state transition, we measured temperature-dependent valence-band spectra using both HAXPES and SXPS. Figure~\ref{fig:valence2}(a) displays the temperature-dependent HAXPES VB spectra recorded with S-polarized light at 300~K and 400~K. A strong reduction of the spectral intensity of feature~$\mathit{A}$ (near binding energy of 0.8~eV)  is observed with increasing temperature, while the rest of the spectrum remains largely unchanged, consistent with previous reports\cite{takegami2023paramagnetic}. Since feature~$\mathit{A}$ is primarily attributed to Co~$3d$ states (as discussed in the previous section), the suppression of its intensity suggests a gradual spin-state crossover between the intermediate- and high-temperature regions \cite{takegami2019valence,barman1994photoelectron}. At low temperatures, Co$^{3+}$ ions in LaCoO$_3$ exist predominantly in the LS ( $t_{2g}^{6}e_{g}^{0}$) configuration, resulting in a valence band dominated by fully occupied $t_{2g}$ states \cite{pandey2008investigation}. With increasing temperature, a fraction of Co$^{3+}$ ions is thermally excited to the HS ( $t_{2g}^{4}e_{g}^{2}$) or IS ($t_{2g}^{5}e_{g}^{1}$ state), leading to partial depopulation of the $t_{2g}$ orbitals and occupation of $e_g$ states. Because the HS state lies only about 0.08~eV above the LS state \cite{korotin1996intermediate}, this energy separation cannot be resolved within the instrumental resolution. Consequently, the observed suppression of the Co~$3d$-derived feature~$\mathit{A}$ with increasing temperature primarily reflects a reduction of $t_{2g}$-derived spectral weight accompanied by a redistribution toward $e_g$ states. This observation suggests that the effective photo-ionization cross section of the $t_{2g}$ orbitals is larger than that of the $e_g$ orbitals, leading to a net decrease in the Co~$3d$ intensity across the spin-state transition. This also indicates that the temperature evolution of the VB spectra is governed by a complex interplay between orbital occupancy changes and polarization dependent photoemission matrix elements. Since Co$^{3+}$ ions are predominantly in the LS ($t_{2g}^{6}e_{g}^{0}$) configuration at room temperature, the valence-band spectra indicate a dominant contribution from Co 3d states with low-spin–like character. A similar trend is observed in Fig.~\ref{fig:valence2}(b), which shows the temperature-dependent SXPS VB spectra measured with Al~K$\alpha$ radiation. Here, the gradual suppression of the Co~3$d$ peak intensity with temperature again reflects the redistribution of spectra weight among $3d$ orbitals associated with spin state transition. The consistent observation of this spectral evolution across two photon-energy regimes, spanning from surface- to bulk-sensitive measurements, confirms that it originates from the intrinsic electronic behavior of LaCoO$_3$ and establishes the spin-state transition as a robust bulk phenomenon. While the spectra in Fig.~\ref{fig:valence2}(b) exhibit only modest changes between 300 and 403 K, clear spectral variations appear when comparing low-temperature (163 K) and higher-temperature measurements. This behavior indicates that LaCoO$_3$ remains predominantly in the low-spin state at low temperatures, with a gradual increase in the HS population setting in above room temperature. Such a temperature-dependent evolution toward an inhomogeneous LS--HS mixture is in excellent agreement with previous reports~\cite{takegami2023paramagnetic,haverkort2006spin,abbate1993electronic,barman1994photoelectron}. This interpretation is further supported by the analysis of the Co~2$p$ core-level spectra presented in the following section.

\subsection{Co 2p Core Level Photo-Emission Spectra}
\begin{figure}[t]
    \centering
    \includegraphics[width=0.8\columnwidth]{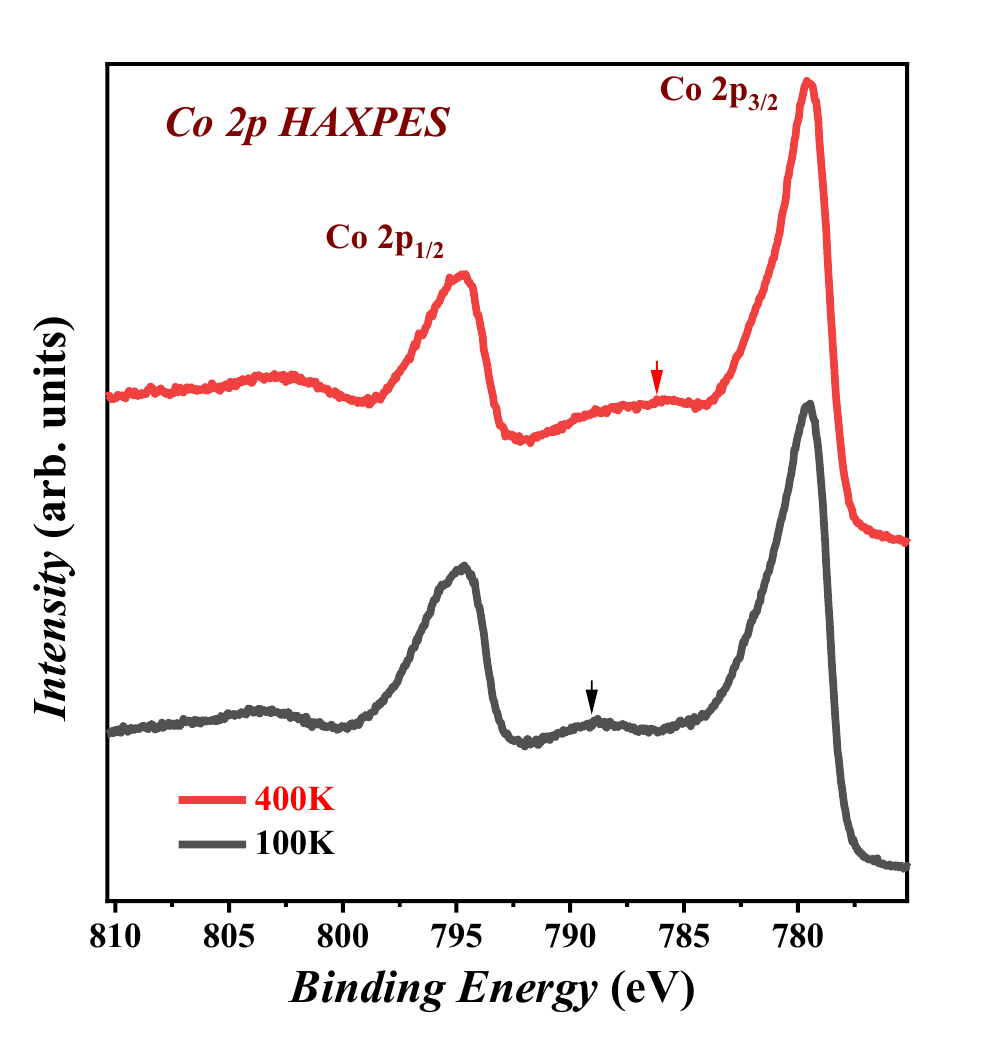}
    \caption{Co~2$p$ core-level HAXPES spectra of LaCoO$_3$ measured at 100 and 400~K.  The emergence of new features around 786~eV (indicated by an arrow) at 400~K and around 789~eV (indicated by an arrow) at 100~K signifies the transition from the LS to the HS state.
    Spectra are vertically offset for clarity.}
    \label{fig:co2p_temp_dep}
\end{figure}

 In Fig.\ref{fig:co2p_temp_dep}, we present the HAXPES of the Co 2$p$ core levels of LaCoO$_3$ recorded at two different temperatures. Core-level (Co 2$p$) photoemission spectra show signatures of spin-state transitions as their main-line and satellite features directly reflect changes in the 3$d$ occupancy and ligand hybridization strength. The high bulk sensitivity of HAXPES makes it ideal for such core-level analysis. The core-hole spin–orbit coupling splits the spectrum into the Co 2$p_{3/2}$ main line near ~780 eV and the Co 2$p_{1/2}$ component at ~795 eV. Apart from these main peaks, weak charge-transfer satellite features appear at higher binding energies. At low temperature (100 K), such a feature—more prominent in this case—is observed around~789 eV\cite{barman1994photoelectron}. The spectra also exhibit  pronounced changes with elevated temperature. As the temperature increases, the main line becomes broader and more asymmetric and an additional feature emerges near 786~eV binding energy marked with an arrow in Fig. \ref{fig:co2p_temp_dep}, which is a clear fingerprint of a thermally induced spin-state transition from LS to HS configurations \cite{barman1994photoelectron}. These features correspond to on-site electronic transitions of the form
\begin{equation}
2p^6 3d^6 \rightarrow 2p^5 3d^{n+1},
\end{equation}
\begin{figure}[htbp]
    \centering
    \includegraphics[width=0.45\textwidth]{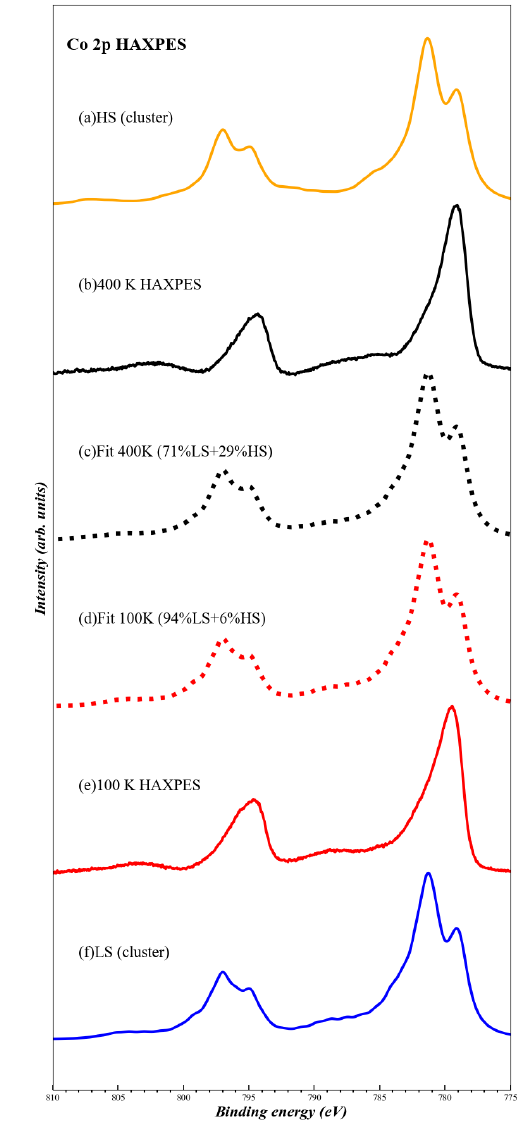}
   \caption{
Comparison between experimental and simulated Co~2$p$ core-level photoemission spectra of LaCoO$_3$.
Panels (a) and (f) show the full-multiplet configuration-interaction cluster calculations for the HS and LS states, respectively.
Panels (b) and (e) display the Co~2$p$ HAXPES experimental spectra measured at 400~K and 100~K.
Panels (c) and (d) present the corresponding theoretical simulations at 400~K and 100~K, respectively, obtained as an incoherent weighted superposition of the calculated LS and HS spectra, with the LS/HS ratios indicated in each panel.
The cluster-model simulations are performed for the LS ($t_{2g}^6$) configuration with $10Dq = 0.80$~eV and $\Delta = 0.80$~eV, and for the HS ($t_{2g}^4e_g^2$) configuration with $10Dq = 0.50$~eV and $\Delta = 0.50$~eV.
All spectra are normalized to the total integrated intensity.
}
    \label{fig:cluster_calc}
\end{figure}
  The emergence of these additional features can be understood within the configuration-interaction (CI) cluster-model framework. In this picture, the Co 2$p$ core hole created during photoemission strongly couples to the Co 3$d$–O 2$p$ valence states, giving rise to a set of charge-transfer final states with mixed configurations of $d^6$, $d^7\underline{L}$, and $d^8\underline{L}^2$ ($\underline{L}$ denotes a ligand hole). In the LS state, strong $p$–$d$ covalency enhances the charge-transferred components ($d^7\underline{L}$, $d^8\underline{L}^2$), whereas in HS state, increased $e_g$ occupancy weakens hybridization, thereby increasing the ionic $d^6$ weight. This redistribution of configuration weights reflects in the feature around 786~eV and 789~eV in the measured spectra, marking the LS–HS transition.
To quantitatively interpret the temperature-driven evolution of the Co~2$p$ HAXPES spectra, we simulated theoretical spectra using a configuration-interaction (CI) cluster model that incorporates full atomic multiplet interactions and hybridization between Co~3$d$ and O~2$p$ orbitals. 
Figure~\ref{fig:cluster_calc} (a) and (f) show the calculated spectra for the HS and LS states, respectively. The simulations reproduce the experimental line shapes well, with our calculated spectra exhibiting a feature near 789~eV for the LS configuration (comparable to the 100~K data) and another near 786~eV for the HS configuration (comparable to the 400~K data), mirroring the experimental trends and remaining in good agreement with previous theoretical works~\cite{abbate1993electronic,takegami2023paramagnetic}. To further examine the possible role of the IS configuration, we compared the experimental Co 2$p$ spectra at 100 K and 400 K with cluster-model simulations corresponding to the IS state. The IS templates were systematically shifted and broadened to match the experimental energy resolution , yet none reproduced either the satellite position or the asymmetric line shape of the Co 2$p_{3/2}$ feature.
This indicates that the IS configuration contributes negligibly to the measured spectra, consistent with previous studies \cite{tomiyasu2017coulomb,asai1994neutron,radaelli2002structural}etc. However the temperature evolution of the Co 2$p$ spectra can be described within the two-state (LS–HS) scenario adopted in our configuration-interaction analysis.The parameters employed in the cluster calculations were optimized for the best agreement with experiment:
$U_{dd} = 5.5$~ eV, $U_{pd} = 7.5$ ~eV, and crystal-field splitting values of $10Dq = 0.80$ eV and $\Delta = 0.80$ ~eV for the LS configuration, and $10Dq = 0.50$~ eV and $\Delta = 0.50$ ~eV for the HS configuration. These values are also consistent with prior reports \cite{takegami2023paramagnetic,haverkort2006spin}.
To determine the temperature-dependent spin-state populations, the experimental Co~2$p$ spectra were modeled as a weighted linear combination of the calculated LS and HS multiplet spectra, as shown in figure~\ref{fig:cluster_calc}. Each calculated spectrum was interpolated onto the experimental energy grid and peak normalized to the Co~2$p_{3/2}$ maximum before fitting.

At 100~K, the best fit corresponds to 94\% LS and 6\% HS contributions, while at 400~K  the LS fractions decrease to 71\% (see figure~\ref{fig:cluster_calc}(c) and (d)). 
The monotonic increase of the HS component with temperature quantitatively confirms the thermally driven spin-state transition in LaCoO$_3$. Overall, the excellent agreement between experiment and simulation demonstrates that the Co~2$p$ HAXPES spectra can reliably quantify the LS--HS crossover in LaCoO$_3$. The derived spin-state evolution---predominantly LS at low temperatures and a mixed, inhomogeneous LS--HS configuration at high temperatures---is fully consistent with the valence-band PES results and previous reports ~\cite{takegami2023paramagnetic,haverkort2006spin,abbate1993electronic,barman1994photoelectron}

\section{Conclusions}

We have investigated the electronic structure and temperature-induced spin-state crossover in LaCoO$_3$ using photoemission spectroscopy with different photon energies, polarizations, and temperatures. The valence-band spectra reveal clear changes in the Co-$3d$ derived states with increasing temperature, indicating a redistribution of spectral weight between $t_{2g}$ and $e_g$ orbitals across the spin-state transition. The bulk-sensitive Co $2p$ HAXPES measurements, supported by cluster calculations, show that LaCoO$_3$ evolves from a low-spin state at low temperatures toward a mixed low-spin and high-spin state at elevated temperatures. The results demonstrate that photoemission spectroscopy provides a direct and reliable probe of spin-state crossover in LaCoO$_3$ and highlight the importance of orbital-selective effects in understanding its electronic structure.

\balance
\section*{ACKNOWLEDGMENTS}

We acknowledge support from the Science and Engineering Research Board (SERB), Government of India, under Grant No.~CRG/2020/003108. Portions of this research were carried out at the PETRA-III light source, DESY. Financial support from the Department of Science and Technology (DST), Government of India, within the framework of the India@DESY collaboration, is gratefully acknowledged.

\bibliographystyle{apsrev4-2}
\bibliography{references}

\end{document}